\documentclass[conference]{IEEEtran}
\IEEEoverridecommandlockouts
\usepackage{amsmath,amsfonts}
\usepackage{algorithmic}
\usepackage{algorithm, color}
\usepackage{array}
\usepackage[caption=false,font=normalsize,labelfont=sf,textfont=sf]{subfig}
\usepackage{textcomp}
\usepackage{stfloats}
\usepackage{url}
\usepackage{verbatim}
\usepackage{graphicx}
\usepackage{cite}
\usepackage[T1]{fontenc}
\usepackage{theorem}
\usepackage{amssymb}

\newtheorem{lemma}{Lemma}\theoremheaderfont{\normalfont\bfseries}

\begin{document}

\title{Integrated Sensing and Communication Beamforming Design Based on Mutual Information}
\author{{Jin Li, Nan Liu} \\

The National Mobile Communications Research Laboratory, Southeast University, Nanjing, China \\

\{lijin,nanliu\}@seu.edu.cn 
\thanks{This work is partially supported by the National Key Research and Development Project under Grants 2019YFE0123600, the European Union’s Horizon 2020 research and innovation programme under the Marie Skłodowska-Curie grant agreement No 872172 (TESTBED2 project: www.testbed2.org), and the Research Fund of National Mobile Communications Research Laboratory, Southeast University (No. 2022A03). }}
%
%

\maketitle

\begin{abstract}
	
Integrated sensing and communication (ISAC) unifies radar sensing and communication, and improves the efficiency of the spectrum, energy and hardware.
In this paper, we investigate the ISAC beamforming design in a downlink system with a communication user and a point radar target to be perceived.
The design criterion for radar sensing is the mutual information between the target response matrix and the echo signals, while the single user transmission rate is used as the performance metric of the communication.
Two scenarios without and with interference from the communication user are investigated. 
A closed-form solution with low complexity and a solution based on the semidefinite relaxation (SDR) method are provided to solve these two problems, respectively.
Numerical results demonstrate that, compared to the results 
obtained by SDR based on the CVX toolbox, the closed-form solution is tight in the feasible region. In addition, we show that the SDR method obtains an optimal solution.
\end{abstract}

\begin{IEEEkeywords}
Integrated sensing and communication, beamforming, mutual information, semidefinite relaxation 
\end{IEEEkeywords}

\section{Introduction}
With the rapid increase of the number of wireless equipments, the existing spectrum is getting more and more crowded.  Thus, it is important to seek ways to make use of the spectrum band of higher frequency or reuse the existing spectral resources. Currently, radar occupies plentiful spectral band, and if wireless communications can make good use of the radar spectrum, it will allieviate the pressure of limited spectrum resources for wireless communications. Hence, integrated sensing and communication (ISAC) is a promising technology to unify radar systems and wireless communication systems. 
As a paradigm of ISAC, dual-functional radar-communication (DFRC) systems use the same signal for both radar and
communication, which greatly improves the efficiency
of the spectrum, hardware and energy compared with the radar and communication co-existence (RCC) systems \cite{overview}, \cite{fantwc}. 

With the deployment of multiple-input multiple-output (MIMO), the DFRC 
base station (BS) can fully exploit the spatial degree of freedoms (DoFs) to enhance the performance of communication and radar sensing \cite{fantcom}. ISAC beamforming design aims at finding a tradeoff between the performance of communication and radar sensing. 
In general, the performance metric of communication is the transmission rate. But there are many performance metrics evaluating radar sensing.
 Recently, minimizing the squared error between the ideal  and practical beampattern under the signal-to-interference-noise ratio (SINR) constraint for communication and transmit power constraint of the BS was proposed in \cite{fantwc}. It shows that with beamforming design, DFRC BS brings a large performance gain over RCC systems. The authors of \cite{CRB} optimized the Cramér-Rao bound (CRB)  for the MIMO DFRC BS, while at the same time achieved a significant performance gain in terms of the beampattern metric compared to \cite{fantwc}. In addition, the CRB was proved to be equal to the mean squared error (MSE) in the case of sensing extended targets \cite{CRB}, \cite{MSE}. 
 In \cite{radarsinr}, the authors maximized the SINR under constant and similarity constraint. It indicates that a better radar performance can be obtained.  
 
 Among performance metrics of radar sensing, mutual information is also a very useful design criterion. It has been proved in \cite{mi} that a high mutual information between the target response matrix and the receiving echo signals implies that radar can achieve accurate classification and estimation performance. The authors of \cite{miandmse} have shown that under the same transmit power constraint,  the performance of MIMO radar waveforms obtained by optimizing the mutual information or the minimum mean squared error (MMSE) are the same when the target impulse vector follows the Gaussian distribution. 
 In the RCC system, designing waveforms by maximizing the mutual information is beneficial to the co-existence of the MIMO radar and the communication in spectrally crowded environments \cite{tangboradar}. For waveform design of adaptive distributed MIMO radar, optimizing the mutual information could bring better performance of target detection and feature extraction \cite{adaptiveradar}. Due to these benefits, this work, for the first time,  adopts the mutual information between the target response and the receiving echo signals as the design criterion of radar
 sensing in ISAC systems.

More specifically, in this paper, we consider a MIMO DFRC BS that works as a colocated MIMO radar with compact antenna spacing to sense a target and communicates with a single user.
 In addition, we compare two cases of radar sensing of a point target. One is without interference from the communication user, and the other is
the case where  the communication user is far away from the BS so that the interference can be regarded as a point-like target.   
The mutual information is adopted as the performance metric of radar sensing while the transmission rate is the design criterion of the communication. A closed-form solution with low complexity is found for the case without interference. Then, we use a series of mathematical transformations to transform the case with interference into a problem solved by the semidefinite relaxation (SDR) method. Finally, numerical results demonstrate the tightness of the closed-form solution in the feasible region, the comparison of beampatterns between these two cases, and the benefits from the spatial DoFs of the MIMO.

\section{System Model}

As shown in Fig. 1, we consider a DFRC system, wherein the MIMO BS, equipped with $N_T$ transmit antennas and $N_R$ receive antennas, communicates with one single-antenna user and senses a target simultaneously.
Let $\mathbf{s}\in \mathbb{C}^{L \times 1}$ be the 
unit-power
data stream to be transmitted to the communication user with $L$ being the length of the communication frame.
For simplicity, we assume that all entries of $\mathbf{s}$ are i.i.d and follow the Gaussian distribution with zero mean and unit variance. By using the law of large numbers, when $L$ is sufficiently large, we have \cite{CRB}
\begin{align}
	\frac{1}{L} \mathbf{s}^H \mathbf{s}  \approx 1.   \label{approx}
\end{align}	
Let vector $\mathbf{w} \in \mathbb{C}^{N_T \times 1}$ represent the dual-functional precoder at the BS, then the transmit DFRC signal  is given by
\begin{align}
	\mathbf{X} = \mathbf{w} \mathbf{s}^H.
\end{align} 
\subsection{Communication Model}
In terms of the communication function, the received signal at the downlink communication user can be written as follows 
\begin{align}
	\mathbf{y}_C^H  =  \mathbf{h}^H \mathbf{X} + \mathbf{n}^H,
\end{align} 
where $\mathbf{h} \in  \mathbb{C}^{N_T \times 1} $  denotes the channel vector between the BS and the communication user, and 
$\mathbf{n} \in \mathbb{C}^{L \times 1}$ denotes the additive white Gaussian noise (AWGN) vector, i.e., $\mathbf{n}  \sim \mathcal{CN} (\mathbf{0},\sigma_{N}^2 \mathbf{I}_{L})$, i.e., each entry of $\mathbf{n}$ is i.i.d and follows the Gaussian distribution with zero mean and  variance $\sigma_N^2$. 

\subsection{Radar Model}
As for the radar function, the BS works as a colocated MIMO radar with uniform linear arrays 
(ULA) so that the angle of departural (AoD) and the angle of arrival (AoA) of each echo are the same.
%
 We assume that a scatterer can be regarded as a point target if it is far away from the DFRC BS  \cite{tangboradar}, \cite{colocatedradar}. Then, the corresponding target response matrix can be modeled by 
\begin{align}
	\mathbf{G}_R =\alpha \mathbf{b}(\theta) \mathbf{a}^{H}(\theta), \label{point}
\end{align}
where $\alpha$ represents the reflection coefficient, $\theta$ is the azimuth angle of the target relative to the BS, and 
$ \mathbf{a}(\theta) \in \mathbb{C}^{N_T \times 1} $ and $\mathbf{b}(\theta) \in \mathbb{C}^{N_R \times 1}$ are the transmit array steering vector and receive array steering vector, respectively. 
\begin{figure}
	\centering \includegraphics[width=2.5in,height=1.7in]{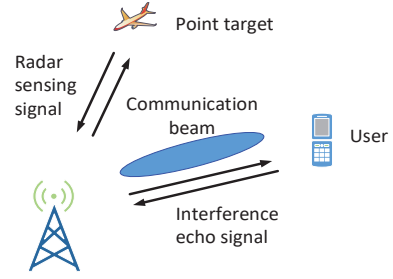}
	\caption{Integrated sensing and communication system model.}
	\label{system} 
\end{figure}

In ISAC, the DFRC BS transmits the integrated signals and
receives the reflected signals from both the radar target and the communication user. In contrast to the case where there is only the radar echo signal, 
the downlink communication system may affect the performance of the target estimation or detection of the radar because the receiving signals of the DFRC BS include echo signals from both the radar sensing target and the communication user \cite{comecho}.


Then, the echo signal received at the DFRC BS is given by
\begin{align} 
	\mathbf{Y}_R &=  \mathbf{G}_R \mathbf{X} + \mathbf{G}_C \mathbf{X} + \mathbf{Z}\nonumber\\
	&=  \mathbf{G}_R \mathbf{w} \mathbf{s}^H +  \mathbf{G}_C \mathbf{w} \mathbf{s}^H + \mathbf{Z},
\end{align}
where  $\mathbf{G}_C \in \mathbb{C}^{N_R \times N_T}$ denotes the target response  matrix of the communication user, and $\mathbf{Z} \in \mathbb{C}^{N_R \times L}$ denotes the Gaussian noise, i.e., each entry of $\mathbf{Z}$ is i.i.d and follows the Gaussian distribution with zero mean and  variance $\sigma_Z^2$.
The random matrix $\mathbf{Y}_R \in \mathbb{C}^{N_R \times L}$ is said to have a matrix variate complex Gaussian distribution with mean matrix $\mathbf{\Lambda} \in \mathbb{C}^{N_R \times L}$ and covariance matrix $\mathbf{\Upsilon} \otimes \mathbf{\Pi}$, where $\mathbf{\Upsilon} \in \mathbb{C}^{N_R \times N_R} $ and $\mathbf{\Pi} \in \mathbb{C}^{L \times L} $ are Hermitian matrices, i.e., $\text{vec}  (\mathbf{Y}_R^H) \sim \mathcal{CN}_{N_R L} (\text{vec}(\mathbf{\Lambda}^H), \mathbf{\Upsilon} \otimes \mathbf{\Pi} )$. Therefore, vectorizing the echo signal matrix $\mathbf{Y}_R^H$, we  obtain
\begin{align} 
	\mathbf{y}_R &= \text{vec}  (\mathbf{Y}_R^H) \nonumber \\
	&= \widetilde{\mathbf{S}} \widetilde{\mathbf{W}} \mathbf{g}_R +\widetilde{\mathbf{S}} \widetilde{\mathbf{W}} \mathbf{g}_C +\mathbf{z} \nonumber \\ 
	&= \widetilde{\mathbf{X}}  \mathbf{g}_R +\widetilde{\mathbf{X}}  \mathbf{g}_C +\mathbf{z},
\end{align}
where $\widetilde{\mathbf{X}}=\widetilde{\mathbf{S}} \widetilde{\mathbf{W}} ,  \widetilde{\mathbf{S}}= \mathbf{I}_{N_R} \otimes \mathbf{s} , \widetilde{\mathbf{W}}= \mathbf{I}_{N_R} \otimes \mathbf{W}^H,   \mathbf{g}_R = \text{vec}(\mathbf{G}_R^H), \mathbf{g}_C = \text{vec}(\mathbf{G}_C^H)$ and $\mathbf{z}=\text{vec}(\mathbf{Z}^H)$. We assume that $\mathbf{z} \sim \mathcal{CN} (\mathbf{0},\sigma_{Z}^2 \mathbf{I}_{LN_R}), \mathbf{g}_R \sim \mathcal{CN} (\mathbf{0}, \mathbf{R}_R)$, and $\mathbf{g}_C \sim \mathcal{CN} (\mathbf{0}, \mathbf{R}_C)$.


Note that the covariance matrix $\mathbf{R}_R$ of \eqref{point} is given by
\begin{align}
	\mathbf{R}_R = \mathbb{E} (\mathbf{g}_R \mathbf{g}_R^H) = \beta^2 (\mathbf{b}^*(\theta) \otimes \mathbf{a}(\theta)) (\mathbf{b}^*(\theta) \otimes \mathbf{a}(\theta))^H,  \label{pointcov}
\end{align}
where $\beta^2 = \mathbb{E} (\alpha \alpha^*) $ denotes the echo signal average strength of the point target.


Following $\mathbf{G}_R$ in \eqref{point}, the target response matrix of the interfering echo signal $\mathbf{G}_C$ is written as  
\begin{align}
	\mathbf{G}_C = \alpha_2 \mathbf{b}(\theta_2) \mathbf{a}^{H}(\theta_2),
\end{align}   
where  $\alpha_2, \theta_2$ are the reflection coefficient and the AoA/AoD of the communication user, respectively. Based on this, the covariance matrices of $\mathbf{g}_R$ and $\mathbf{g}_C$ are given by
\begin{align}
	\mathbf{R}_R  
	& = \beta^2 \text{vec} (\mathbf{P})   \text{vec}^H (\mathbf{P}) \label{covR} \\  
	\mathbf{R}_C 
	& = \gamma^2 \text{vec} (\mathbf{Q})   \text{vec}^H (\mathbf{Q}),  \label{covC}
\end{align}
where $\mathbf{P} = \mathbf{a}(\theta_1)   \mathbf{b}^H(\theta_1),  \mathbf{Q} = \mathbf{a}(\theta_2)   \mathbf{b}^H(\theta_2), \beta^2 = \mathbb{E} (\alpha_1 \alpha_1^*) $ and $ \gamma^2 = \mathbb{E} (\alpha_2 \alpha_2^*)  $.

In the DFRC system, we aim to design a precoding vector $\mathbf{w}$ to meet the requirements of both radar sensing and  communication at the same time, wherein their different performance metrics are discussed in the following subsections.
\subsection{Performance Metric of Communication}
We adopt the transmission rate to evaluate the performance of communication. The transmission rate constraint of the communication user can be represented as 
\begin{align}
	R &= \log_2 (1+SNR) \nonumber \\
	&= \log_2 \Big(1+ \frac{{\lvert \mathbf{h}^H \mathbf{w} \rvert}^2}{  \sigma_{N}^2} \Big) \geq r,  \label{rate}
\end{align}
where $r$ is the target rate required by the communication user.

\subsection{Radar Performance Metric}
For radar sensing, we choose the mutual information between the echo signal of the DFRC BS $\mathbf{y}_R$ and the target response matrix $\mathbf{g}_R$ as the performance metric.  The reason is that as Propositions 1 and 2 in \cite{mi}, we can maximize the mutual information between the parameter of interest and the measurement to obtain more information about the object measured and reduce the measurement error. Specifically, the larger the mutual information, the better the radar estimation, detection and classification. In DFRC systems, the BS knows the transmit signal $\mathbf{X}$ so that the mutual information between $\mathbf{y}_R$ and $\mathbf{g}_R$ given $\mathbf{X}$ can be written as \cite{cover}
\begin{align}
	&\quad \mathcal{I} (\mathbf{y}_R ; \mathbf{g}_R \vert \mathbf{X} = \mathbf{w} \mathbf{s}^H ) \nonumber \\
	&= \mathcal{H} (\mathbf{y}_R \vert \mathbf{X} = \mathbf{w} \mathbf{s}^H) -
	\mathcal{H} (\mathbf{y}_R \vert  \mathbf{g}_R, \mathbf{X} = \mathbf{w} \mathbf{s}^H) \nonumber \\
	&= - \int p(\mathbf{y}_R \vert \mathbf{X} = \mathbf{w} \mathbf{s}^H)	\log \; p(\mathbf{y}_R \vert \mathbf{X} = \mathbf{w} \mathbf{s}^H)  \nonumber \\
	&\quad+ \int p(\mathbf{y}_R \vert \mathbf{g}_R, \mathbf{X} = \mathbf{w} \mathbf{s}^H)  \log \; p(\mathbf{y}_R \vert \mathbf{g}_R, \mathbf{X} = \mathbf{w} \mathbf{s}^H) ,
\end{align}
where $\mathcal{H} (\cdot)$ represents the differential entropy and $p(\cdot)$ represents the probability density function (PDF). 
Thus, the mutual information can be expressed as
\begin{align}
	& \quad \mathcal{I} (\mathbf{y}_R ; \mathbf{g}_R \vert \mathbf{X} = \mathbf{w} \mathbf{s}^H )    \nonumber \\
	& = \log \left[  \det(  L \widetilde{\mathbf{W}} (\mathbf{R}_R + \mathbf{R}_C ) \widetilde{\mathbf{W}}^H   + \sigma_{Z}^2 \mathbf{I}_{KN_R}) \right] \nonumber \\
	& \quad - \log \left[  \det( L \widetilde{\mathbf{W}} \mathbf{R}_C \widetilde{\mathbf{W}}^H   + \sigma_{Z}^2 \mathbf{I}_{KN_R})   \right] \label{MI}. 
\end{align}
 The derivations of \eqref{MI} is based on the property of the matrix determinant $\det (\mathbf{I}_m+\mathbf{AB}) = \det (\mathbf{I}_n+\mathbf{BA})$, the property of the Kronecker product $ (\mathbf{AB}) \otimes (\mathbf{CD}) = (\mathbf{A} \otimes \mathbf{C})(\mathbf{B} \otimes \mathbf{D}) $ \cite{zxd}, and \eqref{approx} under the assumption of sufficiently large $L$.
 
%

\section{ beamforming design without interference from communication to radar sensing }
In this section, 
assuming that the average strength of the echo signals from the communication user is very small, and 
we 
do not consider the interference from the reflected signals of the communication user. 
Then, the formulated problem is given by
\begin{subequations} \label{1}
	\begin{align}
		\max_{\mathbf{w}} \quad  &\log \left[  \det(  L \widetilde{\mathbf{W}} \mathbf{R}_R \widetilde{\mathbf{W}}^H   + \sigma_{Z}^2 \mathbf{I}_{N_R}) \right] \label{1a} \\
		\text{s.t.} \quad & {\Vert \mathbf{w} \Vert}^2 \leq P_0 \label{1b} \\
		&  	R \geq r. \label{1c}
	\end{align}
\end{subequations}
Here, \eqref{1b} is the power constraint for the BS with $P_{0}$ being the maximum transmitting power, and (\ref{1c}) follows from (\ref{rate}).

First, from the expression \eqref{rate}, the constraint \eqref{1c} can be simplified to
\begin{align}
	{\lvert \mathbf{h}^H \mathbf{w} \rvert}^2 \geq \Omega, \label{single SINR}
\end{align}
where $ \Omega = (2^r -1) \sigma_{N}^2 $ and $r$ is the rate threshold of communication user .

Then, we can make some transformations for the objective function \eqref{1a}. By ignoring the angle's index, the covariance matrix $\mathbf{R}_R$ in \eqref{covR} can be reformulated as 
\begin{align}
	\mathbf{R}_R  
	& = \beta^2 \text{vec} (\mathbf{P})   \text{vec}^H (\mathbf{P}), \label{P}
\end{align}
where $\mathbf{P} = \mathbf{a}(\theta) \mathbf{b}^H(\theta) $ and we have used the identity $\mathbf{b} \otimes \mathbf{a} =\text{vec} (\mathbf{a} \mathbf{b}^T) $.
Based on this, \eqref{1a} is given by 
	\begin{align}
		&\quad\log \left[  \det(  L \beta^2 \widetilde{\mathbf{W}} \mathbf{R}_R \widetilde{\mathbf{W}}^H   + \sigma_{Z}^2 \mathbf{I}_{N_R}) \right] \nonumber \\ 
		& = \log  (   L \beta^2  \mathbf{w}^H \mathbf{P} \mathbf{P}^H \mathbf{w}    + \sigma_{Z}^2   ) , \label{1-1} 
	\end{align}
where the derivation of \eqref{1-1} follows the fact that $\text{vec}( \mathbf{A} \mathbf{B} \mathbf{C}) = (\mathbf{C}^T \otimes \mathbf{A}) \text{vec}(\mathbf{B})$, 
$ \det(\mathbf{A}) = \det(\mathbf{A}^*)$ for a Hermitian matrix, and the property of the matrix determinant $\det (\mathbf{I}_m+\mathbf{AB}) = \det (\mathbf{I}_n+\mathbf{BA})$ \cite{zxd}.


Substituting $\mathbf{P} = \mathbf{a}(\theta) \mathbf{b}^H(\theta)$ into \eqref{1-1}, we have
\begin{align}
	\log  (   L \beta^2 \zeta \mathbf{w}^H  \mathbf{a}(\theta)  \mathbf{a}^H(\theta)  \mathbf{w}    + \sigma_{Z}^2   ). \label{1-ex}
\end{align}
Notice that $ \zeta= \mathbf{b}^H(\theta)  \mathbf{b}(\theta)$  is a scalar so we can extract it. Then, ignoring the constants in \eqref{1-ex}, we only need to optimize $\mathbf{w}^H  \mathbf{a}(\theta)  \mathbf{a}^H(\theta)  \mathbf{w} $.

Therefore, problem \eqref{1} can be rewritten as
\begin{subequations} \label{1-2}
	\begin{align}
		\max_{\mathbf{w}} \quad  &  \mathbf{w}^H \mathbf{a}(\theta)  \mathbf{a}^H(\theta) \mathbf{w} \label{1-2a} \\
		\text{s.t.} \quad & \mathbf{w}^H \mathbf{w} \leq P_0 \label{1-2b} \\
		&{\lvert \mathbf{h}^H \mathbf{w} \rvert}^2 \geq \Omega. \label{1-2c}
	\end{align}
\end{subequations}
Although problem \eqref{1-2} can be solved by SDR or successive convex approximation (SCA) based on CVX tools, we provide a closed-form optimal solution with much less complexity in the next lemma.
\begin{lemma} 
	The optimal solution to problem \eqref{1-2} is given by
	\begin{align*}
		\begin{split}
			\mathbf{w} = \left \{
			\begin{array}{lr}
				\sqrt{P_0} \frac{\mathbf{a}(\theta) }{\Vert \mathbf{a}(\theta)  \Vert},  & {\lvert \mathbf{h}^H \mathbf{a}(\theta) \rvert}^2 > \frac{ \Omega {\Vert \mathbf{a}(\theta)  \Vert}^2 }{P_0} \\
				\sqrt{P_0} \frac{\mathbf{h} }{\Vert \mathbf{h}  \Vert},     &  \frac{\lvert \mathbf{a}^H(\theta) \mathbf{h} \rvert}{\Vert \mathbf{a}(\theta)  \Vert \Vert \mathbf{h}  \Vert } =  1 \\
				z_1 \frac{\mathbf{h}}{\Vert \mathbf{h} \Vert} + z_2 \frac{\mathbf{a}(\theta)}{\Vert \mathbf{a}(\theta) \Vert},                                 & otherwise
			\end{array}
			\right.
		\end{split}
	\end{align*}
	where $z_1 = \sqrt{P_0} ( \sqrt{t} - u_2 r ) \frac{\mathbf{a}^T \mathbf{h}^*}{| \mathbf{a}^H \mathbf{h} | }, z_2 = \sqrt{\frac{P_0 (1-t) }{1-r^2}}, t =  \frac{\Omega}{P_0 {\Vert \mathbf{h}  \Vert}^2}, r = \frac{{\lvert \mathbf{a}^H(\theta) \mathbf{h} \rvert} }{  {\Vert \mathbf{a}(\theta)  \Vert} {\Vert \mathbf{h}  \Vert} }$,  and $u_2 = \sqrt{\frac{1-t}{1-r^2}}$. \label{lemma1}
\end{lemma}

\textbf{\textit{Proof: }} Please refer to Appendix A.  \hspace{1.5cm} $\blacksquare$

\section{ beamforming design with interference from communication to radar sensing  }
In general, scattering signals reflected from the communication user can also be received by the DFRC BS. It means that the echo signals received by the BS is comprised of reflecting signals from the desired radar target and the interfering communication user.
The communication user is assumed to be far away from the BS so that the scattering signal interference  can be regarded as a point target interference as shown in \eqref{point}.

Similar to \eqref{1}, we would like to maximize the mutual information between the echo signal $\mathbf{y}_R$ and the target response matrix of the radar target $\mathbf{g}_R$ under power and tranmission rate constraint. The optimization problem is formulated as follows
\begin{subequations} \label{2}
	\begin{align}
		\max_{\mathbf{w}} \quad  &\log \left[  \det(  L \widetilde{\mathbf{W}} (\mathbf{R}_R + \mathbf{R}_C ) \widetilde{\mathbf{W}}^H   + \sigma_{Z}^2 \mathbf{I}_{KN_R}) \right]  \nonumber \\
		&- \log \left[  \det( L \widetilde{\mathbf{W}} \mathbf{R}_C \widetilde{\mathbf{W}}^H   + \sigma_{Z}^2 \mathbf{I}_{KN_R})   \right] \label{2a} \\
		\text{s.t.} \quad & {\Vert \mathbf{w} \Vert}^2 \leq P_0 \label{2b} \\
		&  	R \geq r. \label{2c}
	\end{align}
\end{subequations}


Different from the problem in \eqref{1}, the problem in \eqref{2} is difficult to tackle due to the existence of  interference from the communication user in the objective function. In the following, we transform the nonconvex problem \eqref{2} into a semidefinite programming (SDP) problem by a series of mathematical transformations.
First, we transform the nonconvex objective function \eqref{2a} into a convex one. It is observed that \eqref{2a} has a similar expression to \eqref{1a}, which motivates us to tackle \eqref{2a} in the same way as \eqref{1a}. Thus, \eqref{2a} can be rewritten as
\begin{subequations} \label{2-1}
	\begin{align}
		&\quad \log \left[  \det(  L \widetilde{\mathbf{W}} (\mathbf{R}_R + \mathbf{R}_C ) \widetilde{\mathbf{W}}^H   + \sigma_{Z}^2 \mathbf{I}_{KN_R}) \right] \nonumber \\
		&\quad - \log \left[  \det( L \widetilde{\mathbf{W}} \mathbf{R}_C 
		\widetilde{\mathbf{W}}^H   + \sigma_{Z}^2 \mathbf{I}_{KN_R})   \right] \nonumber \\
		& = \log \left[  \det (  L \beta^2 \mathbf{P}^H \mathbf{w} \mathbf{w}^H \mathbf{P}  + L \gamma^2 \mathbf{Q}^H \mathbf{w} \mathbf{w}^H \mathbf{Q} + \sigma_{Z}^2 \mathbf{I}_{N_R} ) \right] \nonumber \\
		&\quad - \log \left[  \det (  L \gamma^2 \mathbf{Q}^H \mathbf{w} \mathbf{w}^H \mathbf{Q} + \sigma_{Z}^2 \mathbf{I}_{N_R} ) \right] \label{2-1a}\\
		& = \log \left[  \det (  L \beta^2 \mathbf{P}^H \mathbf{w} \mathbf{w}^H \mathbf{P}  + L \gamma^2 \mathbf{Q}^H \mathbf{w} \mathbf{w}^H \mathbf{Q} + \sigma_{Z}^2 \mathbf{I}_{N_R} ) \right]  \nonumber \\
		& \quad - \log  (   L \gamma^2 \mathbf{w}^H \mathbf{Q} \mathbf{Q}^H \mathbf{w}  + \sigma_{Z}^2   ), \label{2-1b}
	\end{align}
\end{subequations}
where \eqref{2-1a} follows from similar reasons as \eqref{1-1}.
And the second term in \eqref{2-1b} uses the property of the matrix determinant $\det (\mathbf{I}_m+\mathbf{AB}) = \det (\mathbf{I}_n+\mathbf{BA})$ to discard the determinant. 
The first term in \eqref{2-1b} cannot be simplified by directly using the property of the matrix determinant, 
hence, we need to seek some additional mathematical transformations. Specifically,  by setting parameters 
	\begin{align}
		\mathbf{T} = \left[\begin{array}{cc}
			\mathbf{w} & \mathbf{0} \\
			\mathbf{0} & \mathbf{w}
		\end{array}\right],   \quad 
		\mathbf{M} = \left[ \begin{array}{c} 
			\sqrt{L} \gamma \mathbf{Q} \\
			\sqrt{L} \beta \mathbf{P}
		\end{array}
		\right],
	\end{align}
the first term in \eqref{2-1b} can be recast as follows
\begin{subequations} \label{2-2}
	\begin{align}
		&\quad \log   \det (  L \beta^2 \mathbf{P}^H \mathbf{w} \mathbf{w}^H \mathbf{P}  + L \gamma^2 \mathbf{Q}^H \mathbf{w} \mathbf{w}^H \mathbf{Q} + \sigma_{Z}^2 \mathbf{I}_{N_R} )  \nonumber \\
		& = \log   \det( \mathbf{T}^H  \mathbf{M} \mathbf{M}^H \mathbf{T}  + \sigma_{Z}^2 \mathbf{I}_{2}    )   	\label{2-2b} \\
		& = \log [ ( L \gamma^2 \mathbf{w}^H \mathbf{Q} \mathbf{Q}^H \mathbf{w} + \sigma_{Z}^2 ) ( L \beta^2 \mathbf{w}^H \mathbf{P} \mathbf{P}^H \mathbf{w} + \sigma_{Z}^2 )   \nonumber \\
	   & \quad - L^2 \beta^2 \gamma^2 ( \mathbf{w}^H \mathbf{Q} \mathbf{P}^H \mathbf{w}  )   ( \mathbf{w}^H \mathbf{P} \mathbf{Q}^H \mathbf{w} )   ], \label{2-2d}
	\end{align}
\end{subequations}
where we use $\det (\mathbf{I}_m+\mathbf{AB}) = \det (\mathbf{I}_n+\mathbf{BA})$ in \eqref{2-2b} to transform the matrix inside the determinant into a $2 \times 2$ matrix. This facilitates the calculation of the determinant.

Combining  \eqref{2-1b} and \eqref{2-2d}, the objective function \eqref{2a} is given by 
\begin{align}
	&  \log [ ( L \gamma^2 \text{Tr}( \mathbf{Q} \mathbf{Q}^H \mathbf{w} \mathbf{w}^H) + \sigma_{Z}^2 ) ( L \beta^2 \text{Tr}( \mathbf{P} \mathbf{P}^H \mathbf{w} \mathbf{w}^H) + \sigma_{Z}^2 )   \nonumber \\
	&  - L^2 \beta^2 \gamma^2 \text{Tr}(  \mathbf{Q} \mathbf{P}^H \mathbf{w} \mathbf{w}^H )   \text{Tr}(  \mathbf{P} \mathbf{Q}^H \mathbf{w} \mathbf{w}^H )   ]  \nonumber \\
	& - \log  (   L \gamma^2 \text{Tr}(\mathbf{Q} \mathbf{Q}^H \mathbf{w} \mathbf{w}^H ) + \sigma_{Z}^2   ),  \label{2-3}
\end{align}
where we use the property of the matrix trace, i.e., $\text{Tr}(\mathbf{A}\mathbf{B}\mathbf{C}\mathbf{D}) = \text{Tr}(\mathbf{B}\mathbf{C}\mathbf{D}\mathbf{A})$. 
Then, maximizing \eqref{2-3} is equivalent to maximizing the function in \eqref{2-4}, shown at the top of the next page.
\begin{figure*}[htbp]
	\centering
\begin{align}
	\frac{( L \gamma^2 \text{Tr}( \mathbf{Q} \mathbf{Q}^H \mathbf{w} \mathbf{w}^H) + \sigma_{Z}^2 ) ( L \beta^2 \text{Tr}( \mathbf{P} \mathbf{P}^H \mathbf{w} \mathbf{w}^H) + \sigma_{Z}^2 ) -   L^2 \beta^2 \gamma^2 \text{Tr}(  \mathbf{Q} \mathbf{P}^H \mathbf{w} \mathbf{w}^H )   \text{Tr}(  \mathbf{P} \mathbf{Q}^H \mathbf{w} \mathbf{w}^H )}{L \gamma^2 \text{Tr}(\mathbf{Q} \mathbf{Q}^H \mathbf{w} \mathbf{w}^H ) + \sigma_{Z}^2 }. \label{2-4}
\end{align}
\hrulefill
\begin{subequations} \label{2-5}
	\begin{align}
		\min_{\mathbf{w}, t} \quad  & -t \label{2-5a} \\
		\text{s.t.} \quad & \left[ \begin{array}{cc}
			L \beta^2 \text{Tr}( \mathbf{P} \mathbf{P}^H \mathbf{W}  ) + \sigma_{Z}^2 - t  &
			L \gamma \beta \text{Tr} (\mathbf{Q} \mathbf{P}^H \mathbf{W}  ) \\
			L \beta \gamma \text{Tr}( \mathbf{P} \mathbf{Q}^H \mathbf{W} ) & 
			L \gamma^2 \text{Tr}(\mathbf{Q} \mathbf{Q}^H \mathbf{W} ) + \sigma_{Z}^2
		\end{array}  \right]  \succeq 0 \label{2-5b}.
	\end{align}
\end{subequations}
\hrulefill
\end{figure*}

The objective function in \eqref{2-4} is still difficult to tackle so we transform it into an SDP optimization problem by means of the Schur complement with an auxiliary variable. Maximizing \eqref{2-4} can be reformulated as problem \eqref{2-5}, shown at the top of the next page, 
where $t$ is an auxiliary variable and $\mathbf{W} = \mathbf{w} \mathbf{w}^H$. 

After the series of  mathematical transformations above, we have converted the maximization of \eqref{2a} into an SDP optimization problem, i.e., \eqref{2-5}. Next, we discuss the transformations for constraints \eqref{2b} and \eqref{2c}. With the equation $\mathbf{W} = \mathbf{w} \mathbf{w}^H$, \eqref{2b} can be rewritten as 
\begin{align}
	\mathbf{w}^H \mathbf{w} = \text{Tr}( \mathbf{W} ) \leq P_0 . \label{2-6}
\end{align}
Based on \eqref{single SINR}, \eqref{2c} can be expressed as
\begin{align}
	\mathbf{h}^H \mathbf{w} \mathbf{w}^H \mathbf{h} = \text{Tr}( \mathbf{h} \mathbf{h}^H \mathbf{W} ) \geq \Omega. \label{2-7}
\end{align} 
With \eqref{2-5}, \eqref{2-6} and \eqref{2-7}, the optimization problem in \eqref{2} can be recast as
\begin{subequations} \label{2-8}
	\begin{align}
		\min_{\mathbf{W}, t} \quad  & -t \label{2-8a} \\
         \text{s.t.} \quad &  \eqref{2-5b}	\label{2-8b} \\
		& \text{Tr}( \mathbf{W} ) \leq P_0, \label{2-8c} \\
		& \text{Tr}( \mathbf{h} \mathbf{h}^H \mathbf{W} ) \geq \Omega, \label{2-8d} \\
		& \mathbf{W} \succeq \mathbf{0}, \text{rank}(\mathbf{W}) = 1. \label{2-8e}
	\end{align}
\end{subequations}
By applying the SDR technique to drop the nonconvex constraint $\text{rank}(\mathbf{W}) = 1$, problem \eqref{2-8} can be relaxed into a standard SDP problem and solved by using the CVX toolbox. The feasible rank-one solution to problem \eqref{2} can be found based on the Gaussian randomization technique \cite{sdr}.

\section{Numerical Results}
In this section, numerical results are provided to evaluate the proposed algorithms. 
The DFRC BS is set to have $N_T = N_R = 6$ antennas.
The communication channel is an i.i.d. Rayleigh fading channel, where each entry has a complex Gaussian distribution with zero mean and unit variance, i.e., $\mathbf{h}  \sim \mathcal{CN} (\mathbf{0},\mathbf{I}_{N_T}) $. 
 The rate threshold is $r = 6 \,\text{bps/Hz} $ and the length of the communication frame is $L = 30$. The maximum transmitting power is $P_0 = 40  \,\text{dBm}$ and the noise powers are set as $ \sigma_{N}^2 = 20 \,  \text{dBm} $ and $\sigma_{Z}^2 = 30 \,\text{dBm}$. The transmit and receive array steering vectors of the ULA at the DFRC BS are respectively given by
\begin{align}
	\mathbf{a}(\theta) = [ 1, e^{-i \frac{2\pi d_1 }{\lambda} \sin \theta }, \dots, e^{-i \frac{2\pi (N_T - 1)d_1 }{\lambda} \sin \theta }  ],   \nonumber \\
	\mathbf{b}(\theta) = [ 1, e^{-i \frac{2\pi d_2 }{\lambda} \sin \theta }, \dots, e^{-i \frac{2\pi (N_R - 1)d_2 }{\lambda} \sin \theta }  ], \nonumber
\end{align}
where $\lambda$ is the wavelength, and $d_1$, $d_2$ denotes the transmit and receive antenna separations, respectively. Here, we consider $ d_1 = d_2 = \frac{\lambda}{2}$ and the average strength of the echo signals  of the point targets are $ \beta =1, \gamma = 100$. We assume that the point radar target is located at $0^{\circ}$ and  the interference of the point target of the communication user is located at $30^{\circ}$. 
For convenience, 
``W/O-inter'' represents the scheme without communication interference solved by problem \eqref{1} and ``W-inter'' denotes the scheme with interference addressed by problem \eqref{2}.

 Table \uppercase\expandafter{\romannumeral1} compares the computational time between the closed-form solution and the SDR method for problem \eqref{1}. It can be seen that the computational time of the closed-form solution is much faster than that of the SDR method. This phenomenon explains that the proposed closed-form solution has a lower complexity compared to the conventional SDR method based on the Gaussian randomization. 
\begin{table}[htb]   
	\begin{center}   
		\caption{The comparision of computational time  between the closed-form solution and the SDR method for problem \eqref{1}.}  
		\label{table1} 
		\begin{tabular}{|c|c|c|}   
			\hline   \textbf{Algorithms} & \textbf{Run-Time (s)}    \\
			\hline   W/O-inter based on the closed-form solution & 0.009  \\ 
			\hline   W/O-inter based on the SDR method & 0.62   \\      
			\hline   
		\end{tabular}   
	\end{center}   
\end{table}

Fig. 2 depicts the beampattern of the radar target without and with interference from the communication user. 
The beampatterns of the interference scheme and the non-interference scheme almost coincide at the radar target location of $0^{\circ}$, with a slight gap of about 0.05 dB. By contrast, the beampattern of the interference scheme is suppressed at $-30^{\circ}$. This can be explained as follows:  the angle of interference is far away from the radar target, and thus the interference to the radar sensing can be efficiently suppressed by the proposed algorithm at the desired radar target location.
\begin{figure}[htbp]
	\centering \includegraphics[width=3in,height=2.2in]{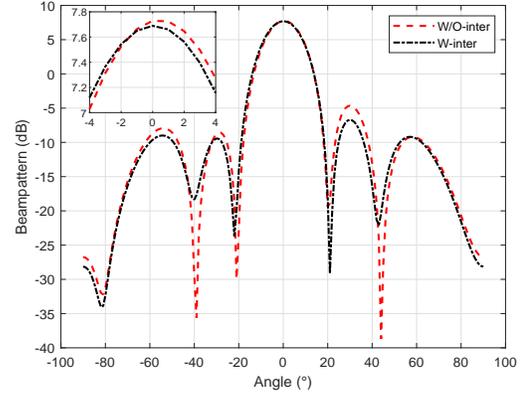}
	\caption{Beampattern comparision of problem \eqref{1} and \eqref{2}.}
	\label{beampattern} 
\end{figure}


Fig. 3 plots the mutual information as a function of the maximum transmitting power for different numbers of transmit and receive antennas. 
 Obviously, a higher mutual information can be attained by a larger transmit power, which leads to a more accurate radar sensing performance. Then, the interference of the point target has very little effect on the performance of radar sensing.
 It is of interest that the mutual information gap between W/O-inter and W-inter schemes is reduced for a large antenna number of $N_T = N_R =12$. This is because the DFRC BS can obtain more spatial DoFs from the increase of the number of antennas, which is beneficial to radar sensing.
 \begin{figure}[htbp]
 	\centering \includegraphics[width=3in,height=2.2in]{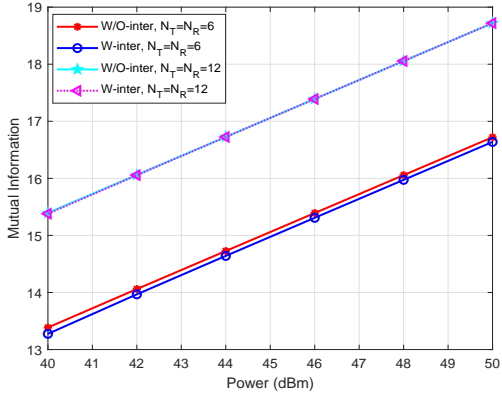}
 	\caption{Mutual information versus maximum transmitting power, when $N_T = N_R =6$ and $N_T = N_R =12$.}
 	\label{power-mi} 
 \end{figure}

\section{Conclusion}
This paper studied the beamforming design of the DFRC BS maximizing the performance metric of the mutual information between the target response and the receiving echo signals. We investigated two cases: one is radar sensing without interference, and the other is radar sensing with interference from the point target of the communication user. For the case of radar sensing without interference, a closed-form solution is proposed, which achieves almost the same performance as the CVX toolbox, but with much lower complexity. For the case of radar sensing with interference, an SDR method is proposed that obtains the optimal solution.

\begin{appendices}        
	\section{ The Proof OF Lemma 1  }        
The power constraint \eqref{1-2b} must be satisfied with equality for the optimal solution $\mathbf{\hat{w}}$. We prove this by contradiction. 
Assume that the optimal solution is $\mathbf{w}_1$ with the power $\mathbf{w}_1^H \mathbf{w}_1 = P_1 < P_0 $. Then, there exists a feasible solution $\mathbf{w}_2 = \sqrt{\frac{P_0}{P_1}} \mathbf{w}_1$ satisfying the constraints \eqref{1-2b} and \eqref{1-2c}, i.e.,
\begin{align}
	&\mathbf{w}_2^H \mathbf{w}_2 = \frac{P_0}{P_1} \mathbf{w}_1^H \mathbf{w}_1 = P_0, \nonumber \\
	&{\lvert \mathbf{h}^H \mathbf{w}_2 \rvert}^2 = \frac{P_0}{P_1}{\lvert \mathbf{h}^H \mathbf{w}_1 \rvert}^2 > {\lvert \mathbf{h}^H \mathbf{w}_1 \rvert}^2 \geq \Omega.
\end{align}
Note that the feasible solution $\mathbf{w}_2$ has a larger objective value than $\mathbf{w}_1$, while $\mathbf{w}_2$ satisfies the constraints \eqref{1-2b} and \eqref{1-2c}. Therefore, the optimal solution $\mathbf{\hat{w}}$ must satisfy the power constraint \eqref{1-2b} with equality. 

Hence, we can reformulate the problem \eqref{1-2} as
\begin{subequations}\label{app1}
	\begin{align} 
		\max_{\mathbf{w}} \quad  &  \mathbf{w}^H \mathbf{a}(\theta)  \mathbf{a}^H(\theta) \mathbf{w}  \label{app1a} \\
		\text{s.t.} \quad & \mathbf{w}^H \mathbf{w} = P_0  \label{app1b} \\
		&{\lvert \mathbf{h}^H \mathbf{w} \rvert}^2 \geq \Omega. \label{app1c}
	\end{align}
\end{subequations}

Then, we discuss the two cases of SINR constraints \eqref{app1c}. First, when the SINR constraint \eqref{app1c} is inactive, i.e, ${\lvert \mathbf{h}^H \mathbf{w} \rvert}^2 > \Omega$,  we only need to consider the power constraint \eqref{app1b}. It can be easily verified that the optimal solution to \eqref{app1} is $\mathbf{\hat{w}} = \sqrt{P_0} \frac{\mathbf{a}(\theta)}{\Vert \mathbf{a}(\theta) \Vert}$.

Second, when the SINR constraint \eqref{app1c} is active, i.e., \eqref{app1c} becomes an equality, we normalize the parameters in the problem \eqref{app1} by taking $\mathbf{c} = \frac{\mathbf{w}}{\Vert \mathbf{w}  \Vert} , \mathbf{v}_2 = \frac{\mathbf{a}(\theta)}{\Vert \mathbf{a}(\theta)  \Vert}  , \mathbf{v}_1 = \frac{\mathbf{h}}{\Vert \mathbf{h}  \Vert}$, and $t = \frac{\Omega}{P_0 {\Vert \mathbf{h}  \Vert}^2}$. The problem can be reformulated as 
\begin{subequations} \label{app3}
	\begin{align}
		\max_{\mathbf{c}} \quad  &  \mathbf{c}^H \mathbf{v}_2  \mathbf{v}_2^H \mathbf{c}  \label{app3a}\\
		\text{s.t.} \quad & \mathbf{c}^H \mathbf{c} = 1  \label{app3b} \\
		& \mathbf{c}^H \mathbf{v}_1  \mathbf{v}_1^H \mathbf{c}  = t  \label{app3c},
	\end{align}
\end{subequations}
 From Cauchy-Schwarz inequality, we obtain $ 0 < t = \frac{{\lvert \mathbf{h}^H \mathbf{w} \rvert}^2 }{ {\Vert \mathbf{h}  \Vert}^2 {\Vert \mathbf{w}  \Vert}^2 }\leq 1 $.

By using Lemma 2 in \cite{proof1} and $\mathbf{c} = \frac{\mathbf{w}}{\Vert \mathbf{w}  \Vert}$, the optimal solution $ \mathbf{\hat{w}}$ is given by
	\begin{align*}
	\begin{split}
		\mathbf{\hat{w}} = \left \{
		\begin{array}{lr}
		 \sqrt{P_0} \frac{\mathbf{h} }{\Vert \mathbf{h}  \Vert},     &  \frac{\lvert \mathbf{a}^H(\theta) \mathbf{h} \rvert}{\Vert \mathbf{a}(\theta)  \Vert \Vert \mathbf{h}  \Vert } =  1 \\
			z_1 \frac{\mathbf{h}}{\Vert \mathbf{h} \Vert} + z_2 \frac{\mathbf{a}(\theta)}{\Vert \mathbf{a}(\theta) \Vert},                               &  0 \leq \frac{\lvert \mathbf{a}^H(\theta) \mathbf{h} \rvert}{\Vert \mathbf{a}(\theta)  \Vert \Vert \mathbf{h}  \Vert } <  1
		\end{array}
		\right.
	\end{split}
\end{align*}
	where $z_1 = \sqrt{P_0} ( \sqrt{t} - u_2 r ) \frac{\mathbf{a}^T \mathbf{h}^*}{| \mathbf{a}^H \mathbf{h} | }, z_2 = \sqrt{\frac{P_0 (1-t) }{1-r^2}}, t =  \frac{\Omega}{P_0 {\Vert \mathbf{h}  \Vert}^2}, r = \frac{{\lvert \mathbf{a}^H(\theta) \mathbf{h} \rvert} }{  {\Vert \mathbf{a}(\theta)  \Vert} {\Vert \mathbf{h}  \Vert} }$,  and $u_2 = \sqrt{\frac{1-t}{1-r^2}}$.

Hence, the proof is complete.

\end{appendices}

\bibliographystyle{IEEEtran}
\bibliography{conference}

\end{document}